\documentclass[12pt,preprint2]{aastex}
\received{--}
\revised{--}
\accepted{--}
\cpright{}{2008}
\slugcomment{To be submitted to ApJ}
\shorttitle{SECCHI-COR2 CME Mass}
\shortauthors{Colaninno \& Vourlidas}
\begin{document}
\title{First Determination of the True Mass of Coronal Mass Ejections: A Novel Approach to Using the Two \textsl{STEREO\/} Viewpoints}
\author{Robin C. Colaninno}
\affil{George Mason University, Fairfax, VA 22030, USA}  
\email{robin.colaninno@nrl.navy.mil}

\author{Angelos Vourlidas}
\affil{Code 7663, Naval Research Laboratory, Washington, DC 20375, USA}
\email{vourlidas@nrl.navy.mil}
\begin{abstract}
The twin Sun Earth Connection Coronal and Heliospheric Investigation (SECCHI) COR2 coronagraphs of the Solar Terrestrial Relations Observatory (\textsl{STEREO\/})  provide images of the solar corona from two view points in the solar system. Since their launch in late 2006, the \textsl{STEREO\/} Ahead (A) and Behind (B) spacecrafts have been slowly separating from Earth at a rate of 22.5 degrees per year. By the end of 2007, the two spacecraft were separated by more than 40 degrees from each other. At this time, we began to see large-scale differences in the morphology and total intensity between coronal mass ejections (CMEs) observed with SECCHI-COR2 on \textsl{STEREO\/}-A and B. Due to the effects of the Thomson scattering geometry, the intensity of an observed CME is dependent on the angle it makes with the observed plane of the sky. From the intensity images, we can calculate the integrated line of sight electron density and mass. We demonstrate that is is possible to simultaneously derive the direction and true total mass of the CME if we make the simple assumption that the same mass should be observed in COR2-A and B.
\end{abstract}

\section{INTRODUCTION}


Coronal mass ejections (CMEs) have been extensively studied and their
general properties are well known after a complete solar cycle of
observations with the Large Angle and Spectrometric Coronagraphs
\citep[LASCO;][]{brueckner}. There exists a large body of literature
detailing the speed, width, and position angle of individual events as
well as statistics for larger samples \citep{stcyr00,
  yashiro04}. There are fewer studies on CME mass and
consequently on their kinetic energy. \citet{vou00, vou02} and
\citet{subra07} published statistics on the mass and energies of LASCO
CMEs and described their analysis methods. 

Because the CME observations are the projection of the three dimensional erupting
structure on the plane of the sky, the measured (width, height, brightness)
and derived (speed, mass, energy) quantities are also projected on the plane
and represent lower limits of the true, un-projected CME
properties. The projection effects on these quantities can be estimated by making
assumptions about the CME propagation direction and shape, but the true
three dimensional properties of the CME remains difficult to estimate reliably \citep{vrsnak07}.

In the case of CME total mass, \citet{vou00} showed that CME masses
are underestimated by about a factor of two, for most cases. This
estimation was supported by three dimensional magnetohydrodynamics
model calculations by \citet{lugaz05}. But the models are idealized
representations of the CME structure and are subject to many
assumptions, leaving some doubts about the fidelity of mass and
kinetic energy measurements. Multiple viewpoint observations of CMEs
offer the best way so far to derive their true properties and quantify the
validity of the projected CME measurements.

The twin Sun Earth Connection Coronal and Heliospheric Investigation (SECCHI) COR2 coronagraphs \citep{howard} of the Solar Terrestrial Relations Observatory \citep[\textsl{STEREO\/};][]{kaiser} provide such observations.  Since their launch in late 2006, the \textsl{STEREO\/} Ahead (A) and Behind (B) spacecraft have separated from Earth at a rate of $22.5^\circ$ per year. By the end of 2007, the two spacecraft were separated from each other by more than $40^\circ$. At that time, we began to see large-scale differences in both the morphology and total intensity between the same CMEs observed with SECCHI-COR2 on \textsl{STEREO\/}-A and B.

The differences in the CME morphology seen by SECCHI are the result of projecting the complex optically thin structure of the CME through the different lines of sight of the COR2-A and B coronagraphs. However, the differences in the total intensity of the CME are mostly due to the different Thomson scattering geometry through the CME plasma. It has long been established that the visible emission of the K-corona originates from the scattering of photospheric light by the coronal electrons \citep{minnaert, hulst, billings} via the Thomson scattering mechanism \citep{jackson}. The scattering strength for a given electron depends on the angle $\chi$, between the vector from the electron to the observer and the radius from the electron to the center of the Sun and the distance from the electron to the Sun. Along any line of sight (LOS), the maximum emission at a fixed radial distance occurs at the point $\chi = 90^\circ$. Within the field of view of a coronagraph, the maximum emission is approximately along a plane. This plane is referred to as the plane of sky (POS) of the observer. Away from the POS, the scattering efficiency decreases. The angle along the LOS away from the POS is $\theta$. Thus the observed intensity of a CME is dependent on the angle, $\theta$, its electrons make with the POS. From the intensity images, we can calculate the electron density and mass for various values of $\theta$. Historically, mass estimates have been calculated for the $\theta=0$ condition, which is the minimum value of the mass. Corrections for this conditions where $\theta>0$, increases the true mass.  

A goal of the \textsl{STEREO\/} mission are to determine true properties of CMEs, including their propagation direction. Ultimately, these goals can be achieved by full three dimensional reconstruction of the CMEs. In this paper, we present a novel way to use the two viewpoints of \textsl{STEREO\/} to locate the CME in longitude. We simply require that the total mass of a CME be the same when the mass calculation is corrected for the two viewpoints. We further demonstrate that in doing so it is possible to simultaneously derive the direction and true total mass of the CME. In \S~2, we begin by calibrating our mass calculations by comparing the total mass measurements from \textsl{SOHO\/}-LASCO and SECCHI. In \S~3, we describe in detail our method for estimating the direction and total mass of the CME using two viewpoints. In this section, we also present the results for eight CMEs observed in COR2. In \S~3.2, we give an expression for the dependence of the de-projected CME mass with height. Finally, in \S~4, we present a discussion of our method and conclusions of our work in this paper.
\section{Mass Analysis Procedure for SECCHI-COR2 Images}

To calculate the total mass of a CME, we first calibrate the images to the customary units of mean solar brightness. We then subtract from the event sequence an
image just prior to the appearance of the CME. This subtraction removes the background F-corona, static K-corona and any residual stray light that has not been removed during the calibration. Thus, we are left with the brightness changes caused by the CME. 

Because we do not know their distribution along the LOS \textit{a priori\/}, we must make the usual assumption that all the electrons are located on the POS. We can then estimate the number of electrons by taking the ratio of the observed brightness ($B_{obs}$) to the brightness of a single electron at a given angular distance, $B_{e}(\theta)$. The brightness, $B_{e}(\theta)$, is calculated analytically from the scattering geometry using the equations in \cite{billings}. To convert the electron density to mass, we assume that the ejected material comprises a mix of completely ionized hydrogen and $10\%$ helium. The mass at each pixel in the image is then calculated using the equation :
\begin{equation}
m = {{B_{obs}} \over {B_{e}(\theta)}} \times 1.97 \times 10^{-24} g.
\label{prime}
\end{equation}
We note that there are two significant advantages of these mass (or electron density) images. First, instrumental effects such as vignetting are removed and secondly the effect of Thomson scattering is removed. Consequently, the image brightness is directly related to the number of electrons along the LOS, regardless of where it is located in the field of view.

\begin{figure*}[p]
\includegraphics[scale=.85]{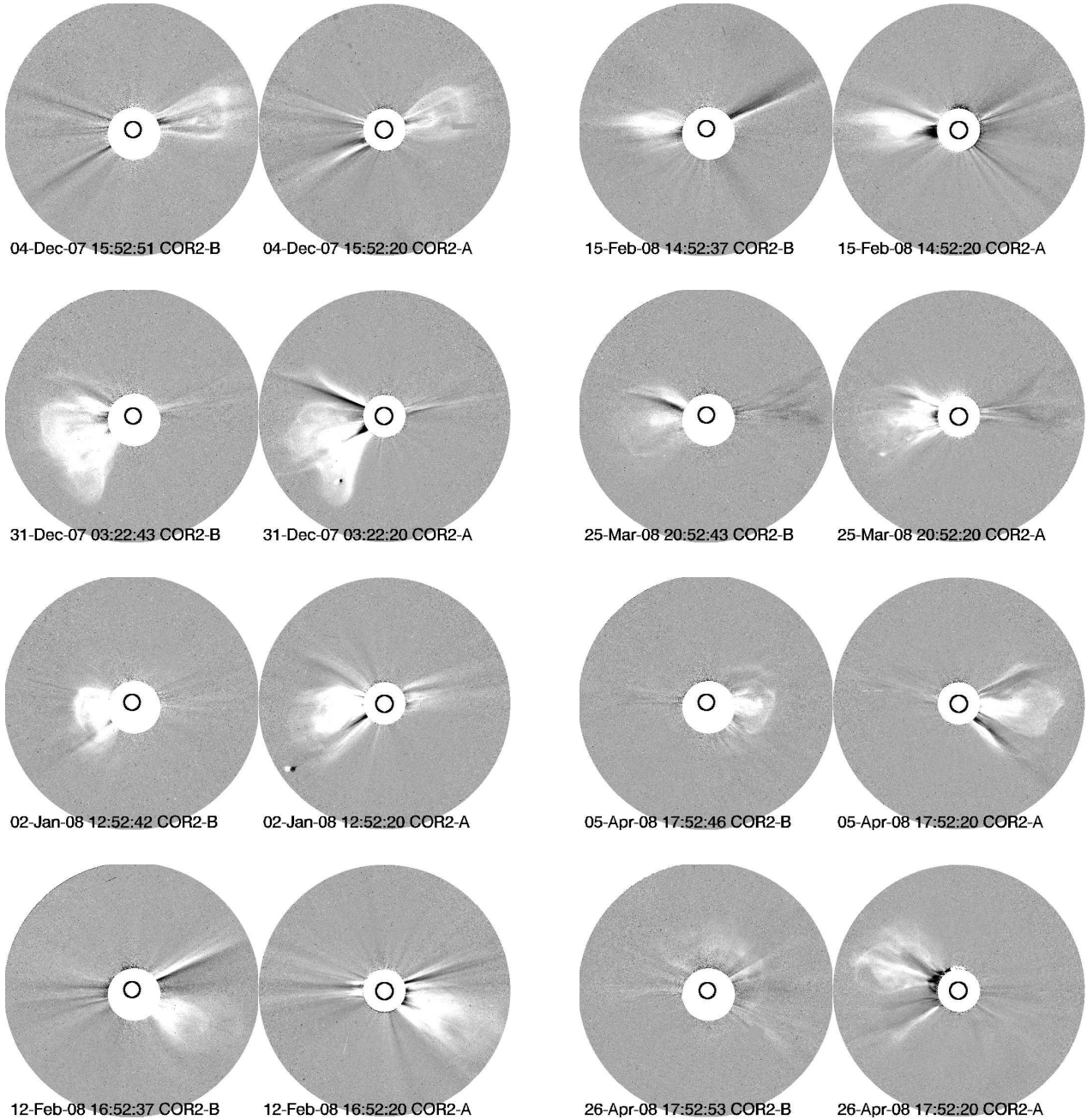}
\caption{Mass images of studied events calculated with $\theta=0$. The images are shown with the same scaling. The left image of each pair is from COR2-B while the right is from COR2-A. The dependence of the CME morphology and total mass on the viewing angle is evident in most events.}
\label{event_figure}
\end{figure*}

Once the brightness value of each pixel in the image is converted to grams, we calculate the total mass by summing the values in the region of the image containing the CME. We perform this procedure for all the images of a time sequence until the leading edge of the CME leaves the COR2 field of view. As an example, Figure~\ref{event_figure} shows the calibrated mass images for the eight CMEs that we studied in COR2-A and B. The dependence of the CME appearance on the viewing angle is evident in most events, especially on the 2008 April 26 event.

\subsection{Cross-Calibration with LASCO Mass Calculations}

To verify our mass analysis procedure for SECCHI-COR2 images, we compare COR2 mass measurements to LASCO C2 and C3 measurements. Validation of the COR2 data is a necessary step since this is the first time that mass measurements from the COR2 instruments have been presented. The availability of concurrent LASCO observations is fortunate for the analysis of SECCHI data since the calibration of the LASCO coronagraphs is very well known \citep{jeff} and CME masses have been studied with LASCO data \citep{vou00,vou02}. For the cross-calibration, we chose events that occurred early in 2007 when the \textsl{STEREO\/} spacecraft were closest to the Sun-Earth line and the \textsl{SOHO\/} spacecraft. The POS is essentially the same for all instruments since the \textsl{STEREO\/} spacecraft were $\leq 2^\circ$ from Earth. In the next section, we will explore the differences in the observed intensities caused by the separation of \textsl{STEREO\/} from Earth.

\begin{figure*}[p]
\includegraphics[scale=.75]{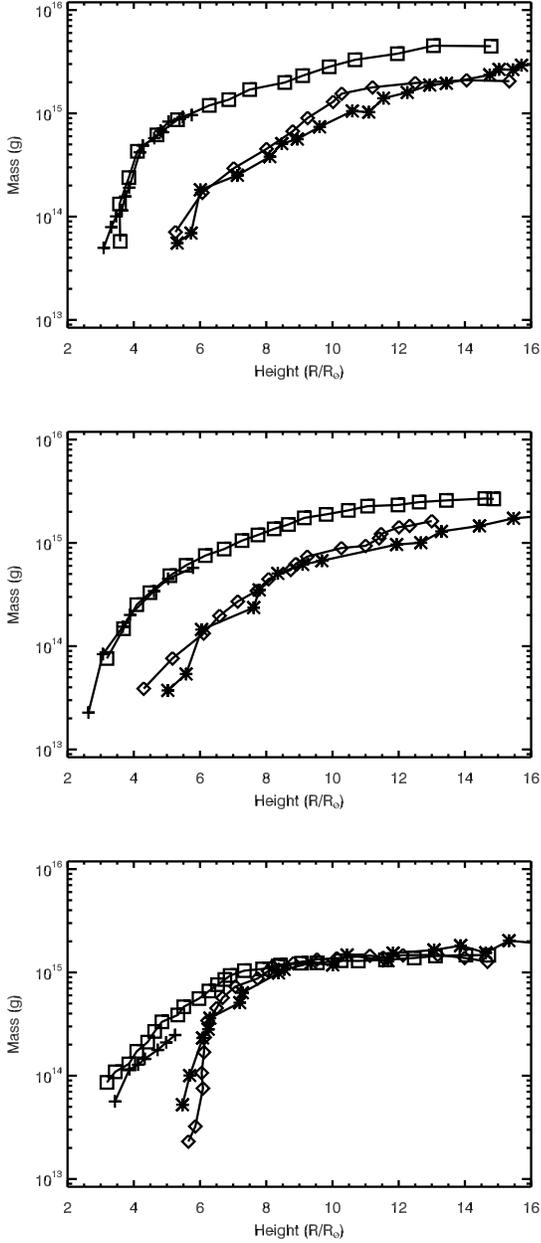}
\caption{Cross-calibration of total mass measurements from LASCO-C2
  (plus), LASCO-C3 (star), SECCHI COR2-A (square) and SECCHI COR2-B
  (diamond) for CMEs observed on 2007 February 9 (top), 2007 March 21
  (middle), and 2007 March 31 (bottom). The good agreement with LASCO C2
and C3 data demonstrates that the COR2 images can be used with
confidence for analysis of CME masses.}
\label{calibration_plot}
\end{figure*}

We calculated the total mass using the procedure describe in the previous section. The results for the four coronagraphs (LASCO C2 and C3, COR2-A and B) are shown in Figure~\ref{calibration_plot}. The LACSO C2 has field of view (FOV) from $\sim 2.5$ to $7 R_{\odot}$  and the LASCO C3 has a FOV from $\sim 4.0$ to $30 R_{\odot}$.  For the cross comparison, we choose to compare the COR2-A data to the LASCO C2 and the COR2-B data to the LASCO C3. We choose to do the comparison in this way because, unfortunately, stray-light in the COR2-B data limits the usable inner FOV and dynamic range early in the mission. Thus, we began the COR2-B measurements at $4.0 R_{\odot}$ for easy of comparison with LASCO C3. In the COR2-A data, we can observe the CME at $2.5 R_{\odot}$ which is comparable to the LASCO C2. Thus we are observing the same area of the CME in both LASCO C2 and C3 with the COR2-A and B, respectively. 

For all three events, the data from the LASCO C2 coronagraph matches well with the data from the COR2-A and the LASCO C3 data matches the data from COR2-B. As the CMEs expands, the difference in the inner FOV has less of an effect on the total mass and the data points converge for the LASCO C3 and COR2-A and B coronagraphs.  The good agreement with LASCO C2 and C3 data demonstrates that the COR2 images can be used with confidence for analysis of CME masses.  

The COR2 mass profiles in Figure~\ref{calibration_plot} provide another important result by verifying that the CME mass increases with height reaching a constant value in the middle corona, above $10 R_{\odot}$ as was suggested by \citet{vou00}. We will further analyze and discuss this behavior in \S~3.2.

\section{A Novel Approach Using the Two \textsl{STEREO\/} Viewpoints}

\begin{figure*}[p]
\includegraphics[scale=.5]{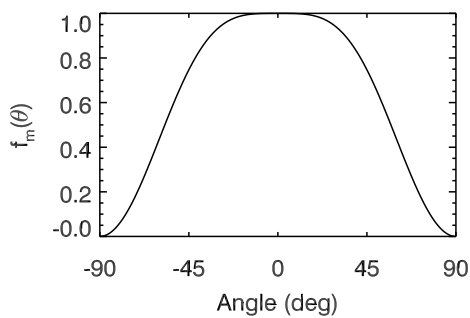}
\caption{The normalized mass function gives the angular dependence of the total brightness of a single scattering electron normalized to the brightness at $\theta=0$. We can use this function to relate the mass calculated using the POS assumption to the true total mass of the CME.}
\label{thom}
\end{figure*}

\begin{figure*}[p]
\includegraphics[scale=.5]{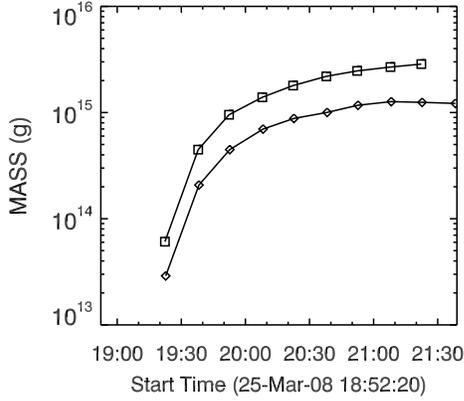}
\caption{March 25, 2008 calculated total mass ($\theta=0$) as a function of time in COR2-A (square) and COR2-B (diamond). The difference in the calculated mass is the result of using an incorrect angle in our Thomson scattering calculation. We will exploit this difference to derive the direction and true total mass of the CME.}
\label{mass_plot1}
\end{figure*}

When simultaneous observations from different viewpoints are available, we can exploit the resulting differences in the mass estimates to obtain not only the true mass but also the direction of the CME. Figure~\ref{mass_plot1} shows the calculated mass versus time for the CME on 2008 March 25 as observed in COR2-A and B for $\theta = 0$. The relationship between the calculated total mass in Figure~\ref{mass_plot1} and the observed total brightness is :
\begin{equation}
M_A = {{B_{A}} \over {B_{e}(\theta =0)}} m_{ej}
\label{eq1}
\end{equation}
\begin{equation}
M_B = {{B_{B}} \over {B_{e}(\theta =0)}} m_{ej}
\label{eq2}
\end{equation}
where again, $B_{e}(\theta)$ is the brightness of a single electron at a given angular distance from the POS and $m_{ej}$ is the mass of the ejected material. For the 2008 March 25 CME, the calculated total mass in COR2-B remains less than the COR2-A mass as the CME expands into the field of view of the coronagraphs. As we have seen previously in Figure~\ref{calibration_plot}, both mass curves converge towards a more or less constant value. Since we are using constant base difference images, we should only be measuring the mass increase caused by the CME. We then conclude that the full extent of the CME is visible in both coronagraphs above $10 R_{\odot}$. Thus we can assume that we are observing the same volume of diffuse material from different angles and we should calculate the same total mass from both COR2-A and B. If this assumption is true, the difference in the calculated masses is a result of using an incorrect angle in our Thomson scattering calculation. The masses calculated in equations~\ref{eq1} and~\ref{eq2} can be expressed as fractions of the true total mass of the CME :
\begin{equation}
M_T f_m(\theta_A) = M_A
\label{MA}
\end{equation}
\begin{equation}
M_T f_m(\theta_B)= M_B
\label{MB}
\end{equation}
where $M_T$ is the true total mass of the CME and $\theta_A$ and
$\theta_B$ are the angular distances of the CME from the POS of COR2-A
and B, respectively. The function, $f_m$, is the ratio of the
brightness of an electron at angle $\theta$ relative to its brightness
on the POS. We will refer to this function as the normalized mass :
\begin{equation}
f_m(\theta) = \frac{B_e(\theta)}{B_e(\theta =0)}.
\label{norm}
\end{equation}
The function, $f_m$, is plotted in figure~\ref{thom}. If the CME were
in the POS of one of the coronagraphs, we would obtain the true total
mass by setting $B_e(\theta =0)$. For CMEs away from the instrument
POS, the calculated mass is some fraction of the true total mass,
expressed by $f_m$. It is of interest to note that if the CME were directed towards one of the coronagraphs then, theoretically, we should not observe any mass.

The angle between the COR2-A and B POS is equal to the \textsl{STEREO\/} spacecraft separation. Thus we can define $\theta_A$ and $\theta_B$, with respect to a common coordinate system. For this coordinate system, the angle, $\theta$, is measured $90^\circ$ from the Sun-Earth line in a right hand coordinate system. Thus equations ~\ref{MA} and ~\ref{MB} can be written as :
\begin{equation}
M_T f_m(\theta + \frac{1}{2}\Delta_{sc}) = M_A
\label{tA}
\end{equation}
\begin{equation}
M_T f_m(\theta - \frac{1}{2}\Delta_{sc}) = M_B
\label{tB}
\end{equation}
where $\Delta_{sc}$ is the angular separation of the two spacecraft,
and $\theta$ is the angle of propagation of the event. Thus the axis of the coordinate system is equal distance from the COR2-A and B POS. We can now equate the difference in the calculated mass in COR2-A and COR2-B to the true total mass. If we combine equations~\ref{tA} and~\ref{tB}, we have :
\begin{equation}
M f_m(\theta + \frac{1}{2}\Delta_{sc}) - M f_m(\theta - \frac{1}{2}\Delta_{sc}) = M_A - M_B.
\label{eq5}
\end{equation}

We can calculate the true total mass by inverting this function to find the longitudinal direction that satisfies equation~\ref{eq5}. The mass difference, $M_A - M_B$, is plotted as a function of longitudinal direction in Figure~\ref{angle_mass} for separation angles of $10^\circ$ to $90^\circ$. The mass difference is the superposition of
two of the functions shown in Figure~\ref{thom} offset by the
spacecraft separation.  The inversion of the function can lead to more
than one solution for a given mass difference. However, some of the
solutions can be eliminated. In Figure~\ref{angle_mass}, the gray part
of the curve shows where the CME would appear on opposite limbs of the
Sun in the two coronagraphs. The dotted part of the curve is where the
CME would appear as a halo in one of the coronagraphs. A simple
inspection of the images would immediately reveal which of the
solutions should be chosen and which eliminated.

As the separation of the spacecraft increases, the range of observable
mass differences also increases. The extrema of the mass difference are related to the normalized mass function by : 
\begin{equation}
(M_A - M_B)^{\ast} = f_m(90^\circ - \Delta_{sc}).
\label{eq6}
\end{equation}
Thus when the separation is $0^\circ$ the extrema is zero and when the separation is $90^\circ$ the extrema are equal to the true total mass. If the difference in our calculated total mass is outside the range of solution for equation ~\ref{eq5}, then we are observing intensity that is not from the CME. An example of this would be instrumental effects or another solar structure, such as a streamer, that was not removed adequately by the base difference.

We applied our method to the eight CMEs shown in
Figure~\ref{event_figure}. For the purposes of comparing the total
mass across the COR2-A and B instruments, we use the same IFOV at $4.0
R_{\odot}$. We selected the largest events observed by COR2 for
spacecraft separation greater than $40^\circ$.  In table~\ref{masstable},
we list the total mass of the CME in COR2 A and B using the POS
assumption ($\theta = 0^\circ$). We then list the true total mass calculated
using the CME direction. The longitudinal direction derived for each CME
with respect to the Sun-Earth line is listed in the next column. For
the majority of the CMEs, the true mass is not significantly different from the
larger of the two masses using the POS assumption. Figure~\ref{thom} shows that the CME mass does not vary significantly between $\pm 20^\circ$ from the POS. The studied CMEs are mainly within $\pm 20^\circ$ of one of the instrument's POS. The spacecraft
separation is given in table~\ref{masstable}.

\subsection{Comparison to Forward Modeling Results}

\begin{figure*}[p]
\includegraphics[scale=.50]{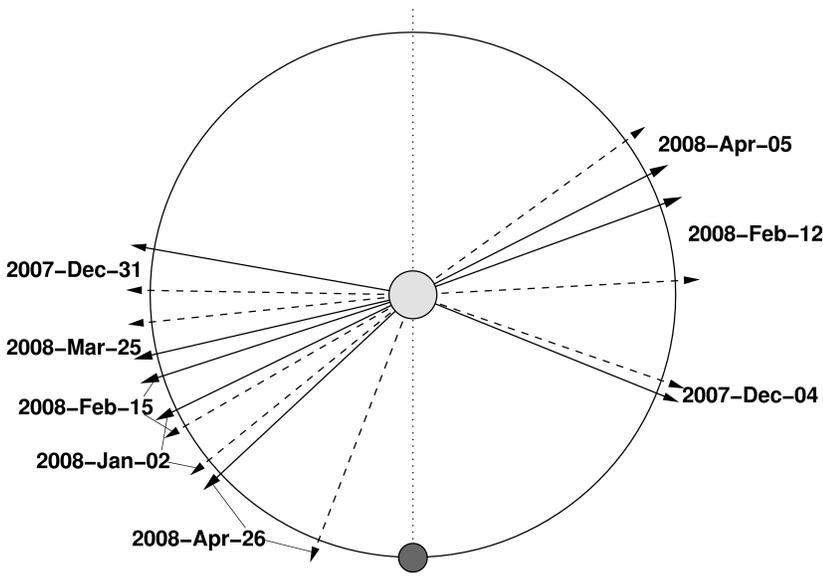}
\caption{Graphical representation of the CME estimated direction for the events in our sample. The dashed lines are the results of our mass method, while the solid lines are obtained from forward modeling \citep{thern08}.} 
\label{polar_plot}
\end{figure*}

As a means of further validating our analysis, we compare our
direction estimates with a completely different approach to estimating
the three dimensional position of the CMEs, namely forward modeling. A
complete description of the forward modeling method can be found in
\citet{thern06}. Briefly, a three dimensional geometric
representation of a flux-rope is fitted to the two spacecraft
views of a CME at a single time. The direction of the CME is taken as
the apex of the flux-rope. The directions from forward modeling for
the CMEs in our sample are given in table~\ref{masstable} \citep{thern08}. Figure~\ref{polar_plot} provides a visual comparison
between the direction results from our mass method and the forward
modeling method. We plot the direction of the CME as calculated using
the mass difference (solid line) and the direction found from the
forward model (dashed line). We have good agreement between all of the
studied CMEs with the exception of the 2008 April 26 event. For this
event, the CME appears as a partial halo in COR2-B which results in a
limitation to the accuracy of our method. A portion of the CME is
behind the occulter and our assumption that we are observing the same
mass in both views is not valid. 

\subsection{CME Mass Variation with Height and Time}

As can be seen in Figures~\ref{calibration_plot} and \ref{mass_plot1},
the total CME mass measurements show a very specific variation with
height and time. Namely, the mass increases rapidly when the CME front
is within $8 R_{\odot}$, and reaches a constant value beyond about
$10 R_{\odot}$. The same behavior was originally seen in LASCO when only a specific, large scale feature of the CME is measured \citep[e.g., the core][]{vou00}. The result for LASCO was treated with caution because the sharpest mass increase occurred in the $5-8 R_{\odot}$ range which is the overlapping region between C2 and C3.

 \citet{vou00} suggests that the mass increase could have been due to instrumental differences between C2 and C3 such as calibration, dynamic range, and resolution. However, the COR2 measurements are taken over an uninterrupted field of view with the same telescope and clearly show that the CME mass increases with time and height then reaches a constant value above about $10 R_{\odot}$.

Therefore, this mass variation with height appears to be a fundamental property of
the ejection process. To further quantify this behavior, we have
fitted the observed mass-height profiles with the analytical function :
\begin{equation}
M(h) = M_c (1 - e^{-h/h_c} ).
\end{equation}
where $M_c$ is the final total mass of the event and $h_c$ is the
height where the mass reaches $~63\%$ of its final mass. The choice of
the function was dictated by the shape of the mass curves. Also this function has the desired behavior of approaching a constant value as the height increases. We have not
explored other functions and we are wary of employing a more complex expression because we do not yet have any theoretical or physical foundation for the variation of CME mass with height. This is an area where CME modelers and theoreticians could provide some useful insight.  

\begin{figure*}[p]
\includegraphics[scale=0.8]{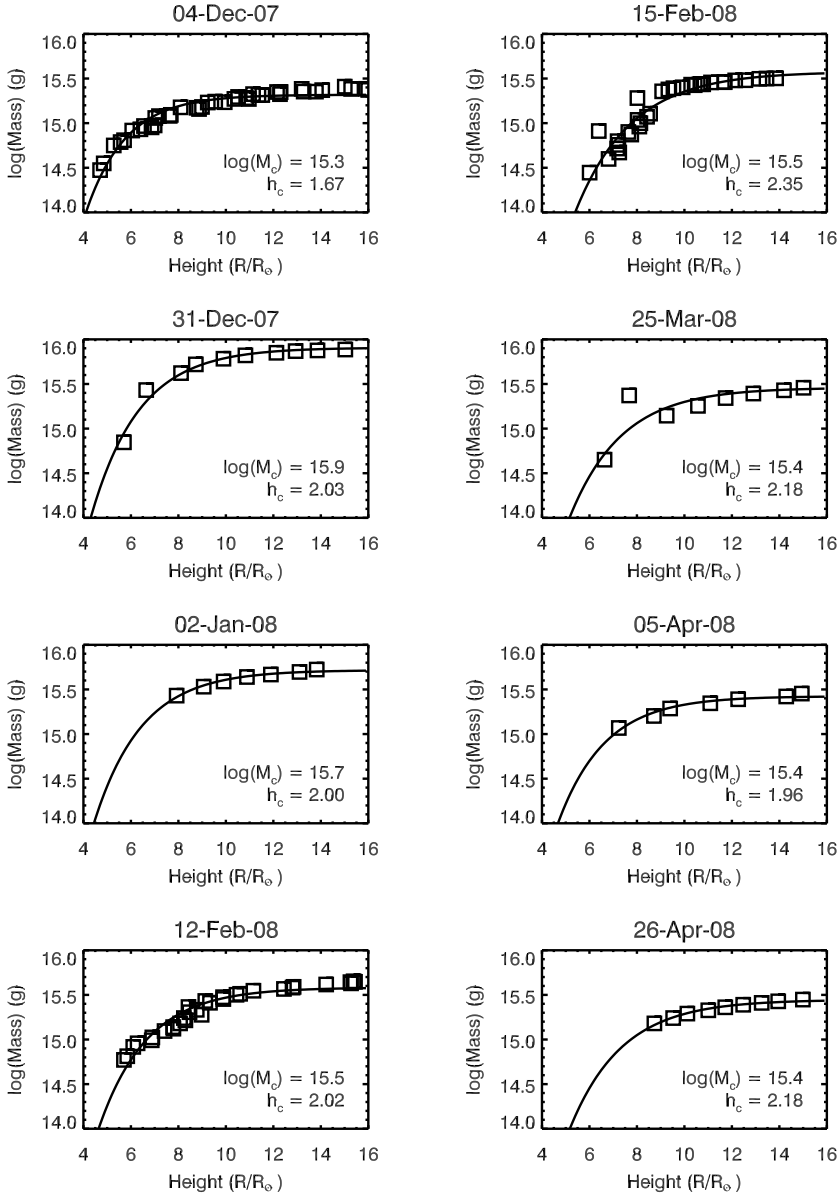}
\caption{The dependence of CME mass on the height of the CME
  front for the eight events in our sample. The fitted final CME mass
  and the scale height for each event are also shown.} 
\label{mass_fit}
\end{figure*}

Before fitting the data, we de-projected both the heights and the CME mass values using
the results in Table~\ref{masstable}. We obtained good fits to all
eight of the events in our paper (Figure~\ref{mass_fit}). For all
events, the scale height ($h_c$) is relatively low in the corona at approximately $2 R_{\odot}$ and $~99 \%$ of the final CME mass is reached by $10 R_{\odot}$. For all of the events in our sample, the final mass is of the order of $10^{15} g$. The variations in the profiles do not seem to correlate
with the speed or the width of the events. It is difficult to reach strong conclusions from our small sample of CMEs taken during a very low period of solar activity. In the future, we plan to investigate the behavior of the CME mass with a larger number of events. However, we are confident that the small
variation in the parameters of the fit suggests that we can adopt
the average profile of the eight events :
\begin{equation}
M(h) = 15.6(1 - e^{-h/2.1})
\end{equation}
 as representative of the mass variation with height for a typical CME. Of course, the mass increase is due to material coming up from below the occulting disk.

\section{DISCUSSION \& CONCLUSIONS}

An implicit assumption in all CME mass calculation methods up to now
has been that the mass of the CME is concentrated into a single plane on the
POS. However, CMEs are three dimensional structures with a
considerable depth along the line of sight. While our two viewpoint
method is an improvement on the POS assumption, $\theta=0$, and results in an
estimate of the CME direction, it still assumes that the CME mass lies
in a plane along that direction. The true width along the LOS remains
unknown. But we can easily estimate the error from this assumption by
calculating the mass ratio between the CME of zero width and CMEs of
various widths. \citet{vou00} showed that this simplification could cause the total mass to be underestimated by up to $15\%$.

To overcome this limitation, we could
use observations at larger heights by combining measurements in SECCHI
HI-1 A and B instruments, for example. Or instead of measuring the
total mass, we could try to measure the mass of the same feature as
long as it can be reliably identified in both COR2 instruments.

There are other factors that could affect the accuracy of the mass
calculation. An obvious one is the noise in the mass images. We
estimate the noise from histograms of empty sky regions. As expected,
the empty sky values are a Gaussian distribution around zero. We define the
error level as one standard deviation of this distribution. The noise levels in the COR2
telescopes are similar and the error is $\sim 9 \times 10^{9}$
g/pixel. This error is comparable to the error in the LASCO mass images
\citep{vou05}. The average per pixel signal in the measured CMEs is
approximately 5 times the noise level and therefore the noise is
insignificant. While the calculation of the CME mass has a low noise
level, the selection of the CME region for the mass calculation can
effect the total mass significantly. In the quiet corona there are
large dense streamers that obscure or interact with the CME. Since we
are using two viewpoints, it is often the case where the streamer
can be isolated from the CME in one view but cannot in another. An
example is the 2008 January 02 event where a streamer is below the
CME in the COR2-A image but bisects it in the COR2-B image. This event
also has the second largest discrepancy with the forward fitting model
($\sim 8^\circ$), so the addition of the streamer may be effecting the
direction finding to some extent. In general, however, it is difficult
to quantify this type of error since it is not always obvious from the
images when a streamer is part of the measured mass. That situation
can be best addressed by simultaneous observations from viewpoints
inside and outside the ecliptic plane.

\begin{figure*}[p]
\includegraphics[angle=90,scale=0.8]{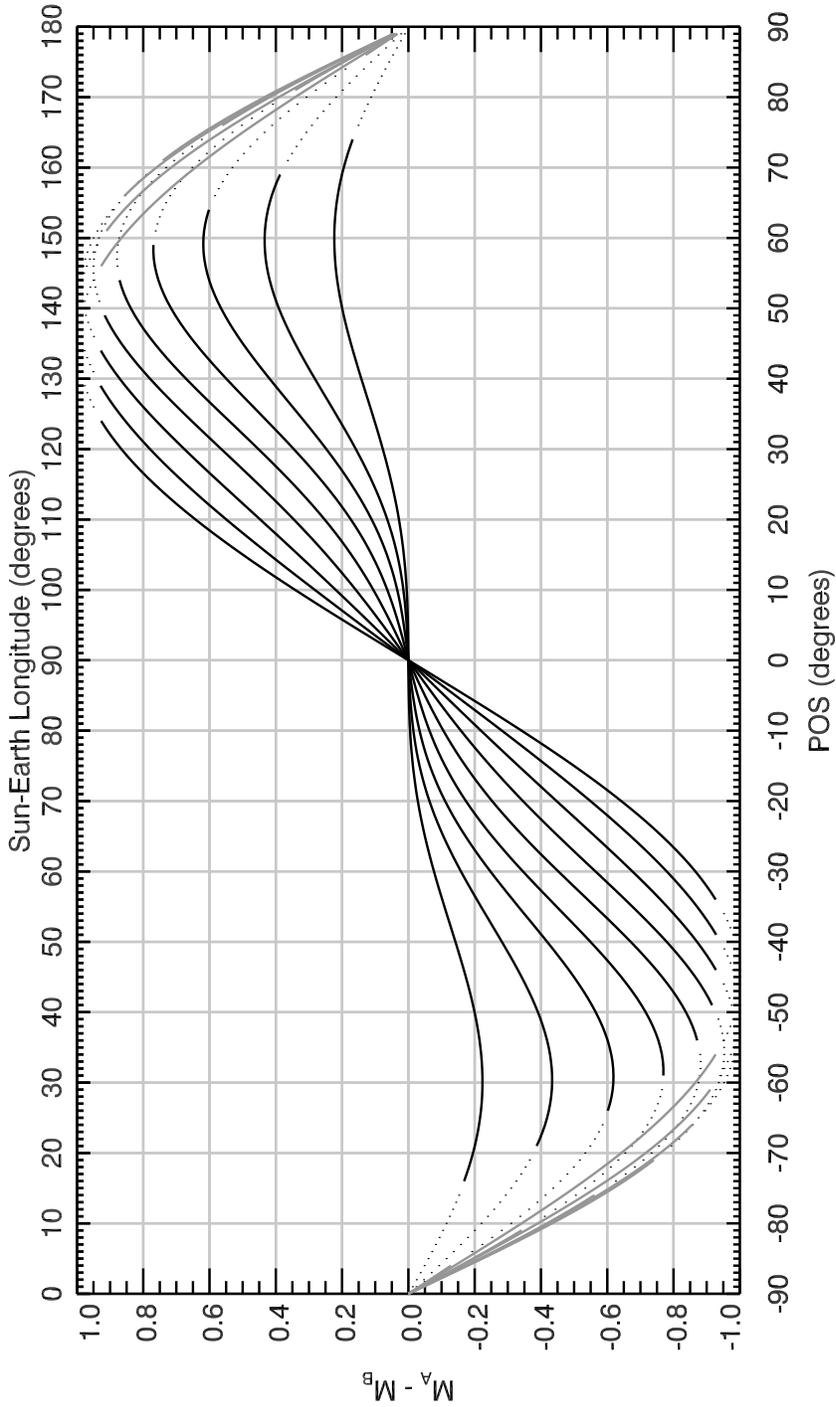}
\caption{Mass difference as a function of direction for spacecraft separations of $10^\circ$ to $90^\circ$ in steps of $10^\circ$. Directions where CMEs would appear as halos (dotted) or on opposite sides of the Sun (gray) can be eliminated as possible solutions. We can calculate the true total mass by inverting this function to find the longitudinal direction for a given mass difference and spacecraft separation.} 
\label{angle_mass}
\end{figure*}

The error in the CME direction arises from the shape of the function
of mass with POS angle (Figure~\ref{angle_mass}). Small changes in the
difference between the two masses can cause large differences in the
direction, for small spacecraft separations. Assuming a typical mass
error estimate of $\sim15\%$, the direction ambiguity becomes
reasonably small ($\lesssim 20^\circ$) for separations larger than
about $50^\circ$.

Another point of discussion is the implication of SECCHI results on the single viewpoint mass measurements of past missions. As we have
mentioned already, all previous work assumed a POS angle of zero for
the CME mass. In table~\ref{masstable}, we show the total mass in each
instrument for $\theta=0$ and the true CME mass. For the majority of
the CMEs, the true mass of the CME is not significantly different from
the larger of the two masses using the POS assumption. In other words,
most of the mass tends to lie near one of the two POS for the
events of our sample. Therefore one has a better chance of observing
the true mass of the CME with two viewpoints for spacecraft separations of $40^\circ-50^\circ$. The lower mass is within a factor of 2 of the true mass for most cases with the exception of
the April 26 event which is a factor of 3 lower. However, this is a
halo event and such discrepancies are expected. Our results validate
the assumptions in \citet{vou00} and the modeling results of
\citet{lugaz05} and suggest that past CME mass measurements are within
a factor of two of the true CME mass, except for halo events.

\acknowledgments We thank A. F. Thernisien for providing the data from
his geometric CME model. The SECCHI data is produced by an international consortium of the NRL, LMSAL and NASA GSFC
(USA), RAL and U. Bham (UK), MPS (Germany), CSL
(Belgium), IOTA and IAS (France).

%
\begin{deluxetable}{c|ccc|cc|ccc}
\tablewidth{34pc}
\tablecaption{CME Direction and Mass}
\tablecolumns{9}
\tablehead{\colhead{CME} & \multicolumn{3}{c}{Mass $10^{15}$g} & \multicolumn{2}{c}{Direction} & \colhead{Separation} & \multicolumn{2}{c}{HEE Lon}\\ & \colhead{B} &\colhead{A} &\colhead{true} & \colhead{mass}& \colhead{model}& &\colhead{B} &\colhead{A}}
\startdata
2007 Dec 04 & 2.57 & 2.23 & 2.57 &  68 &  71 & 42.16 & -21.43 & 20.73 \\
2007 Dec 31 & 7.68 & 7.10 & 7.70 &-100 & -91 & 43.97 & -22.79 & 21.17 \\
2008 Jan 02 & 3.59 & 5.29 & 5.29 & -64 & -51 & 44.07 & -22.88 & 21.20 \\
2008 Feb 12 & 3.05 & 4.49 & 4.49 & 110 &  93 & 45.56 & -23.67 & 21.89 \\
2008 Feb 15 & 2.12 & 3.18 & 3.18 & -72 & -60 & 45.64 & -23.68 & 21.97 \\
2008 Mar 25 & 1.27 & 2.86 & 2.87 & -78 & -84 & 47.17 & -23.69 & 23.48 \\
2008 Apr 05 & 1.89 & 2.84 & 2.84 & 117 & 126 & 47.83 & -23.72 & 24.11 \\
2008 Apr 26 & 0.94 & 2.78 & 2.80 & -48 & -21 & 49.51 & -23.95 & 25.56 \\
\enddata

\label{masstable}
\end{deluxetable}

%


\begin{thebibliography}{99}
\bibitem[Billings(1966)]{billings} Billings, D. E. 1966, A Guide to
  the Solar Corona (New York: Academic Press) 
\bibitem[Brueckner et al(1995)]{brueckner} Brueckner, G.E. et al. 1995, Sol. Phys., 162, 291  
\bibitem[Howard et al.(1985)]{howard85} Howard, R. A. et al. 1985, \jgr, 90, 8173 
\bibitem[Howard et al.(2008)]{howard} Howard, R. et al. 2008, \ssr, 136, 67 
\bibitem[Jackson (1997)]{jackson} Jackson, J. D. 1997, Classical
  Electrodynamics (3rd ed.; New York: Wiley) 
\bibitem[Kaiser et al.(2008)]{kaiser} Kaiser, M. L. et al. 2008, \ssr, 136, 5
\bibitem[Lugaz et al.(2005)]{lugaz05} Lugaz, N. 2005, \apj, 627, 1019
\bibitem[Minnaert (1930)]{minnaert} Minnaert, M. 1930, Z. Astrophys., 1, 209
\bibitem[Morrill et al.(2006)]{jeff} Morrill, J. S. et al. 2006,
  \solphys, 233, 331
\bibitem[St. Cyr et al.(2000)]{stcyr00} St. Cyr, O. C. et al. 2000, \jgr, 105, 18169
\bibitem[Subramanian and Vourlidas(2007)]{subra07} Subramanian, P., \& Vourlidas, A. 2007, A\&A, 467, 685
\bibitem[Thernisien, Howard, and Vourlidas(2006)]{thern06} Thernisien, A. F.,
  Howard, R., \& Vourlidas, A. 2006, \apj, 652, 763 
\bibitem[Thernisien, Vourlidas, and Howard(2009)]{thern08} Thernisien, A. F.,
  Vourlidas, A., \& Howard, R. 2009,  Sol. Phys., in press
\bibitem[van de Hulst(1950)]{hulst} van de Hulst, H. C. 1950,
  Bull. Astron. Inst. Netherlands, 11, 135  
\bibitem[Vourlidas et al.(2000)]{vou00} Vourlidas, A. et al. 2000,
  \apj, 543, 456
\bibitem[Vourlidas et al.(2002)]{vou02} Vourlidas, A. et al 2002, in
  Proc. of the 10th Europ. Sol. Phys. Meet. 'Solar Variability: From
  Core to Outer Frontiers', Prague, Czech Rep., Wilson, A. (ed), ESA
  SP-506, Dec. 2002, p. 91
\bibitem[Vourlidas (2005)]{vou05} Vourlidas, A. in Coronal and Stellar
  Mass Ejections, IAU Symp. Proc. of the IAU 226,  K. Dere,
  J. Wang, and Y. Yan (eds). Cambridge: Cambridge University Press,
  2005., pp.76-76
\bibitem[Vr\v{s}nak et al.(2007)]{vrsnak07} Vr\v{s}nak, B. et al. 2007, A\&A, 469, 339
\bibitem[Yashiro et al.(2004)]{yashiro04} Yashiro, S. et al. 2004, J. Geophys. Res., 109, A07105, 10.1029/2003JA010282 
\end{thebibliography}
\end{document}